\documentclass{appolb}
\usepackage{epsfig}
\begin{document}
\title{Phenomenological test of the large $N_c$ ChPT predictions\\
         for the pseudoscalar mixing parameters%
\thanks{UAB--FT--591 report.
              To appear in the proceedings of the PHOTON2005
              International Conference on the Structure and Interactions of the Photon,
	     31.8--4.9.2005 Warsaw (Poland).}%
}
\author{R. Escribano\thanks{
                        Work partly supported by the Ramon y Cajal program,
                        the Ministerio de Ciencia y Tecnolog\'{\i}a and FEDER, FPA2002-00748EU,
                        and the EU, HPRN-CT-2002-00311, EURIDICE network.}
\address{Grup de F\'{\i}sica Te\`orica and IFAE, 
        Universitat Aut\`onoma de Barcelona,\\
        E-08193 Bellaterra (Barcelona), Spain}%
        }
\maketitle
\begin{abstract}
A phenomenological analysis of various decay processes is performed in order to test
the large $N_c$ Chiral Perturbation Theory predictions for the octet and singlet
pseudoscalar decay constants and mixing angles.
The results obtained hint at a disagreement with the expectations of this theoretical framework
although the statistical significance is still limited.
\end{abstract}

\section{Introduction}
The mixing pattern of the pseudoscalar decay constants associated to the $\eta$-$\eta^\prime$ system
is usually described in terms of a single mixing angle \cite{Ball:1995zv}.
However, this is not the most general mixing scheme since two axial currents,
the eight component of the octet and the singlet,
can couple to the two physical states, the $\eta$ and $\eta^\prime$.
Therefore, a mixing scheme consisting of two mixing angles should be used instead of the simplest one mixing angle description.
The reason for not using the two mixing angle scheme from the beginning has been the absence of a well established framework able to give some insight on the values of the two different angles and their related decay constants.
Now with the advent of large $N_c$ Chiral Perturbation Theory (ChPT) \cite{Kaiser:2000gs},
where the effects of the pseudoscalar singlet are treated in a perturbative way,
the new two mixing angle scheme receives theoretical support.
The aim of this work is to perform an updated phenomenological analysis
of various decay processes using the two mixing angle description of the
$\eta$-$\eta^\prime$ system.
The analysis will serve us to check the predictions of the large $N_c$ ChPT for the 
pseudoscalar mixing parameters (decay constants and mixing angles).

\section{Large $N_c$ ChPT predictions for the mixing parameters}
\label{LargeNcChPT}
The decay constants of the $\eta$-$\eta^\prime$ system in the octet-singlet basis
$f_P^a\ (a=8,0; P=\eta,\eta^\prime)$ are defined as
\begin{equation}
\label{decaycon}
\langle 0|A_\mu^a|P(p)\rangle=i f_P^a p_\mu\ ,
\end{equation}
where $A_\mu^{8,0}$ are the octet and singlet axial-vector currents whose
divergences are
\[
\label{divaxialos}
\begin{array}{c}
\partial^\mu A_\mu^8=\frac{2}{\sqrt{6}}
(m_u\bar u i\gamma_5 u+m_d\bar d i\gamma_5 d-
2m_s\bar s i\gamma_5 s)\ ,\\[2ex]
\partial^\mu A_\mu^0=\frac{2}{\sqrt{3}}
(m_u\bar u i\gamma_5 u+m_d\bar d i\gamma_5 d+m_s\bar s i\gamma_5 s)+
\frac{1}{\sqrt{3}}\frac{3\alpha_s}{4\pi} G_{\mu\nu}^a\tilde G^{a,\mu\nu}\ ,
\end{array}
\]
where $G^a_{\mu\nu}$ is the gluonic field-strength tensor and
$\tilde G^{a,\mu\nu}\equiv\frac{1}{2}\epsilon^{\mu\nu\alpha\beta}G^a_{\alpha\beta}$
its dual.
The divergence of the matrix elements (\ref{decaycon}) are then written as
\begin{equation}
\label{divdecaycon}
\langle 0|\partial^\mu A_\mu^a|P\rangle=f_P^a m_P^2\ ,
\end{equation}
where $m_P$ is the mass of the pseudoscalar meson.

Each of the two mesons $P=\eta, \eta^\prime$ has both octet and singlet components,
$a=8, 0$.
Consequently, Eq.~(\ref{decaycon}) defines \emph{four independent} decay constants.
Following the convention of Refs.~\cite{Leutwyler:1997yr,Kaiser:1998ds}
the decay constants are parameterized in terms of two basic decay constants
$f_{8}, f_{0}$ and two angles $\theta_{8}, \theta_{0}$
\begin{equation}
\label{defdecaybasisos}
\left(
\begin{array}{cc}
f^8_\eta & f^0_\eta \\[1ex]
f^8_{\eta^\prime} & f^0_{\eta^\prime}
\end{array}
\right)
=
\left(
\begin{array}{cc}
f_8 \cos\theta_8 & -f_0 \sin\theta_0 \\[1ex]
f_8 \sin\theta_8 &  f_0 \cos\theta_0
\end{array}
\right)\ .
\end{equation}

Large $N_c$ ChPT is the low energy effective theory of Quantum Chromodynamics (QCD)
based on an extended chiral symmetry group including the effects of the 
$U(1)_A$ anomaly perturbatively and consisting of a simultaneous expansion in powers of
momenta, quark masses and $1/N_c$ of the most general Lagrangian involving
the nonet of pseudoscalar bosons
(the pseudoscalar singlet is considered as the ninth Goldstone boson).

In the octet-singlet basis this theory predicts \cite{Leutwyler:1997yr,Kaiser:1998ds}:
\begin{equation}
\label{mixparthos}
\begin{array}{l}
f_8^2=\frac{4f_K^2-f_\pi^2}{3}\ ,\quad
f_0^2=\frac{2f_K^2+f_\pi^2}{3}+f_\pi^2\Lambda_1\ ,\\[1ex]
f_8 f_0\sin(\theta_8-\theta_0)=-\frac{2\sqrt{2}}{3}(f_K^2-f_\pi^2)\ .
\end{array}
\end{equation}
These expressions are valid at next-to-leading order in the large $N_c$ ChPT
expansion where the octet-singlet pseudoscalar decay constants
can be written in terms of the known $f_\pi$ and $f_K$ decay constants
and the unknown OZI-rule violating parameter $\Lambda_1$.

\section{Experimental input}
\label{expinput}
\subsection{$\eta,\eta^\prime\to\gamma\gamma$}
We will use as constraints the experimental decay widths of 
$(\eta, \eta^\prime)\rightarrow\gamma\gamma$ \cite{Eidelman:2004wy}
\begin{equation}
\label{expwidths}
\begin{array}{c}
\Gamma(\eta\rightarrow\gamma\gamma)=(0.510\pm 0.026)\ \mbox{keV}\ ,\\[1ex]
\Gamma(\eta^\prime\rightarrow\gamma\gamma)=(4.29\pm 0.15)\ \mbox{keV}\ .
\end{array}
\end{equation}
Analogously to the $\pi^0\rightarrow\gamma\gamma$ case,
one assumes that the interpolating fields $\eta$ and $\eta^\prime$ can be
related with the axial-vector currents in the following way:
\begin{equation}
\label{interfields}
\begin{array}{c}
\eta(x)=\frac{1}{m_\eta^2}
\frac{f^0_{\eta^\prime}\partial^\mu A_\mu^8(x)-
      f^8_{\eta^\prime}\partial^\mu A_\mu^0(x)}
{f^0_{\eta^\prime}f^8_\eta-f^8_{\eta^\prime}f^0_\eta}\ ,\quad
\eta^\prime(x)=\frac{1}{m_{\eta^\prime}^2}
\frac{f^0_\eta\partial^\mu A_\mu^8(x)-
      f^8_\eta\partial^\mu A_\mu^0(x)}
{f^0_\eta f^8_{\eta^\prime}-f^8_\eta f^0_{\eta^\prime}}\ .
\end{array}
\end{equation}
This leads to
\begin{equation}
\label{theowidths}
\begin{array}{c}
\Gamma(\eta\rightarrow\gamma\gamma)=
\frac{\alpha^2 m_\eta^3}{96\pi^3}\left(
\frac{f^0_{\eta^\prime}-2\sqrt{2}f^8_{\eta^\prime}}
{f^0_{\eta^\prime}f^8_\eta-f^8_{\eta^\prime}f^0_\eta}\right)^2=
\frac{\alpha^2 m_\eta^3}{96\pi^3}\left(
\frac{c\theta_0/f_8-2\sqrt{2}s\theta_8/f_0}
{c\theta_0 c\theta_8+s\theta_8 s\theta_0}\right)^2\ ,\\[2ex]
\Gamma(\eta^\prime\rightarrow\gamma\gamma)=
\frac{\alpha^2 m_{\eta^\prime}^3}{96\pi^3}\left(
\frac{f^0_\eta-2\sqrt{2}f^8_\eta}
{f^0_\eta f^8_{\eta^\prime}-f^8_\eta f^0_{\eta^\prime}}\right)^2=
\frac{\alpha^2 m_{\eta^\prime}^3}{96\pi^3}\left(
\frac{s\theta_0/f_8+2\sqrt{2}c\theta_8/f_0}
{c\theta_0 c\theta_8+s\theta_8 s\theta_0}\right)^2\ .
\end{array}
\end{equation}

\subsection{$VP\gamma$ form factors}
We will also use as constraints the radiative decays of lowest-lying vector mesons,
$V\rightarrow(\eta,\eta^\prime)\gamma$, 
and of the radiative decays $\eta^\prime\rightarrow V\gamma$, with $V=\rho, \omega, \phi$.
In order to predict their couplings
we follow closely the method presented in Ref.~\cite{Ball:1995zv} where the description of the
light vector meson decays is based on their relation with the $AVV$ triangle anomaly,
$A$ and $V$ being an axial-vector and a vector current respectively.
The approach both includes $SU_f(3)$ breaking effects and fixes the vertex couplings $g_{VP\gamma}$
as explained below.

In that framework, one starts considering the correlation function
\begin{equation}
\label{corrfun}
i\int d^4x e^{iq_1 x}
\langle P(q_1+q_2)|TJ_\mu^{\rm EM}(x)J_\nu^V(0)|0\rangle=
\epsilon_{\mu\nu\alpha\beta}q_1^\alpha q_2^\beta F_{VP\gamma}(q_1^2,q_2^2)\ ,
\end{equation}
where the currents are defined as
\begin{equation}
\label{defcurr}
\begin{array}{l}
J_\mu^{\rm EM}=\frac{2}{3}\bar u\gamma_\mu u-\frac{1}{3}\bar d\gamma_\mu d-
               \frac{1}{3}\bar s\gamma_\mu s\ ,\\[1ex]
J_\mu^{\rho,\omega}=
\frac{1}{\sqrt{2}}(\bar u\gamma_\mu u\mp\bar d\gamma_\mu d)\ ,
\quad J_\mu^\phi=-\bar s\gamma_\mu s\ .
\end{array}
\end{equation}
The form factors values $F_{VP\gamma}(0,0)$ are fixed by the $AVV$ triangle 
anomaly (one $V$ being an electromagnetic current), and are written in terms 
of the pseudoscalar decay constants and the $\phi$-$\omega$ mixing angle 
$\theta_V$ as
\begin{equation}
\label{FVPgamma00}
\begin{array}{l}
F_{\rho\eta\gamma}(0,0)=\frac{\sqrt{3}}{4\pi^2}
\frac{f^0_{\eta^\prime}-\sqrt{2}f^8_{\eta^\prime}}
{f^0_{\eta^\prime}f^8_\eta-f^8_{\eta^\prime}f^0_\eta}\ ,\quad
F_{\rho\eta^\prime\gamma}(0,0)=\frac{\sqrt{3}}{4\pi^2}
\frac{f^0_\eta-\sqrt{2}f^8_\eta}
{f^0_\eta f^8_{\eta^\prime}-f^8_\eta f^0_{\eta^\prime}}\ ,\\[1ex]
F_{\omega\eta\gamma}(0,0)=\frac{1}{2\sqrt{2}\pi^2}
\frac{(c\theta_V-s\theta_V/\sqrt{2})f^0_{\eta^\prime}-
      s\theta_V f^8_{\eta^\prime}}
{f^0_{\eta^\prime}f^8_\eta-f^8_{\eta^\prime}f^0_\eta}\ ,\\[1ex]
F_{\omega\eta^\prime\gamma}(0,0)=\frac{1}{2\sqrt{2}\pi^2}
\frac{(c\theta_V-s\theta_V/\sqrt{2})f^0_\eta-s\theta_V f^8_\eta}
{f^0_\eta f^8_{\eta^\prime}-f^8_\eta f^0_{\eta^\prime}}\ ,\\[1ex]
F_{\phi\eta\gamma}(0,0)=-\frac{1}{2\sqrt{2}\pi^2}
\frac{(s\theta_V+c\theta_V/\sqrt{2})f^0_{\eta^\prime}+
      c\theta_V f^8_{\eta^\prime}}
{f^0_{\eta^\prime}f^8_\eta-f^8_{\eta^\prime}f^0_\eta}\ ,\\[1ex]
F_{\phi\eta^\prime\gamma}(0,0)=-\frac{1}{2\sqrt{2}\pi^2}
\frac{(s\theta_V+c\theta_V/\sqrt{2})f^0_\eta+c\theta_V f^8_\eta}
{f^0_\eta f^8_{\eta^\prime}-f^8_\eta f^0_{\eta^\prime}}\ .\\[1ex]
\end{array}
\end{equation}
Using their analytic properties, we can express these form factors by
dispersion relations in the momentum of the vector current, which are then
saturated with the lowest-lying resonances:
\begin{equation}
\label{VMD}
F_{VP\gamma}(0,0)=\frac{f_V}{m_V}g_{VP\gamma}+\cdots\ ,
\end{equation}
where the dots stand for higher resonances and multiparticle contributions
to the correlation function.
In the following we assume vector meson dominance (VMD) and thus neglect these
contributions.
The $f_V$ are the leptonic decay constants  of the vector mesons and
can be determined from the experimental decay rates of $V\rightarrow e^+e^- $ \cite{Eidelman:2004wy}.
Finally, we introduce the vertex couplings $g_{VP\gamma}$, which are just the
on-shell $V$-$P$ electromagnetic form factors:
\begin{equation}
\label{gVPgamma}
\langle P(p_P)|J_\mu^{\rm EM}|V(p_V,\lambda)\rangle|_{(p_V-p_P)^2=0}=
-g_{VP\gamma}\epsilon_{\mu\nu\alpha\beta} p_P^\nu p_V^\alpha\varepsilon_V^\beta(\lambda)\ ,
\end{equation}
which are measured from the decay widths of $P\rightarrow V\gamma$ and
$V\rightarrow P\gamma$ \cite{Eidelman:2004wy}.
Eq.~(\ref{VMD}) allows us to identify the $g_{VP\gamma}$ couplings defined in
(\ref{gVPgamma}) with the form factors $F_{VP\gamma}(0,0)$ listed in (\ref{FVPgamma00}).
The couplings are expressed in terms of the octet and singlet mixing angles $\theta_8$ and $\theta_0$,
the pseudoscalar decay constants $f_8$ and $f_0$, the $\phi$-$\omega$ mixing angle $\theta_V$,
and the corresponding vector decay constants $f_V$.

\section{Results}
\label{results}
In order to test the large $N_c$ ChPT predictions for the pseudoscalar mixing parameters
we must first know the values of the decay constants and mixing angles
preferred by the experimental data.
We have performed various fits to this set of experimental data assuming the two mixing angle scheme
of the $\eta$-$\eta^\prime$ system.
The theoretical constraint $f_8=1.28 f_\pi$ or $1.34 f_\pi$\footnote{
A value of $f_8=1.34 f_\pi$ is obtained if chiral logs and higher order contributions
are also taken into account \protect\cite{Kaiser:1998ds}.}
is relaxed in order to test the dependence of the result on the value of this parameter.
The experimental constrain $\theta_V=(38.7\pm 0.2)^\circ$ is also relaxed to test the stability
of the fit.
The results are presented in Table \ref{table1}.
\begin{table}
\centering
{\small
\begin{tabular}{ccc}
\hline\hline
\\[-1ex]
Assumptions & Results & $\chi^2/\textrm{d.o.f.}$\\[1ex]\hline
\\[-1ex]
$\theta_8$ and $\theta_0$ free  & $\theta_8=(-22.2\pm 1.4)^\circ$   & 40.5/5\\[1ex]
$f_8=1.28 f_\pi$                            & $\theta_0=(-5.5\pm 2.3)^\circ$     &\\[1ex]
$\theta_V=(38.7\pm 0.2)^\circ$  & $f_0=(1.25\pm 0.04) f_\pi$           &\\[1ex]\hline
\\[-1ex]
$\theta_8$ and $\theta_0$ free  & $\theta_8=(-22.8\pm 1.4)^\circ$   & 30.0/5\\[1ex]
$f_8=1.34 f_\pi$                            & $\theta_0=(-4.7\pm 2.2)^\circ$     &\\[1ex]
$\theta_V=(38.7\pm 0.2)^\circ$  & $f_0=(1.26\pm 0.04) f_\pi$           &\\[1ex]\hline
\\[-1ex]
$\theta_8$ and $\theta_0$ free  & $\theta_8=(-23.8\pm 1.4)^\circ$   & 17.9/4\\[1ex]
$f_8$ free                                       & $\theta_0=(-1.1\pm 2.3)^\circ$     &\\[1ex]
$\theta_V=(38.7\pm 0.2)^\circ$  & $f_8=(1.51\pm 0.05) f_\pi$           &\\[1ex]
                                                         & $f_0=(1.32\pm 0.05) f_\pi$           &\\[1ex]\hline
\\[-1ex]
$\theta_8$ and $\theta_0$ free  & $\theta_8=(-23.7\pm 1.6)^\circ$   & 17.9/3\\[1ex]
$f_8$ free                                       & $\theta_0=(-1.0\pm 2.5)^\circ$     &\\[1ex]
$\theta_V$ free                             & $f_8=(1.51\pm 0.05) f_\pi$           &\\[1ex]
                                                        & $f_0=(1.32\pm 0.05) f_\pi$           &\\[1ex]
                                                        & $\theta_V=(38.5\pm 2.4)^\circ$   &\\[1ex]
\hline\hline
\end{tabular}
}
\caption{Results for the $\eta$-$\eta^\prime$ mixing angles and decay constants
in the octet-singlet basis of the two mixing angle scheme.
The fitted experimental data includes the decay widths of
$(\eta,\eta^\prime)\rightarrow\gamma\gamma$, $V\rightarrow P\gamma$ and $P\rightarrow V\gamma$.}
\label{table1}
\end{table}

As seen from Table \ref{table1},
the $\theta_8$ and $\theta_0$ mixing angle values are different at the $3\sigma$ level.
The fit with $1.34 f_\pi$ is slightly better than that with $f_8=1.28 f_\pi$.
When $f_8$ is left free the experimental data seem to prefer a value higher than the one
predicted by ChPT ($f_8=1.28 f_\pi$),
while for the parameters $\theta_8, \theta_0$ and $f_0$ our values are in agreement with
those of Refs.~\cite{Leutwyler:1997yr,Feldmann:1999uf}.
This increase is translated into a considerably better fit.
However, the analysis of the parameter correlation coefficients reveals a strong positive
correlation between $f_8$ and $\theta_0$ ($+0.547$)
---the latter quantity being taken as a negative number in the present convention;
in other terms, the correlation to the \emph{absolute} value of $\theta_0$ is negative.
The correlations between $f_8$ and $\theta_8$ or $\theta_8$ and $\theta_0$ are negative and
much weaker ($-0.159$ and $-0.100$, respectively).
It is also observed that when $f_8$ is fixed, the remaining correlation coefficients are smaller.
Finally, it is seen in Table \ref{table1} that relaxing the experimental constrain
$\theta_V=(38.7\pm 0.2)^\circ$ does not produce any effect on the fit.
A numerical prediction for all the observables used in the fits can be found in
Ref.~\cite{Escribano:2005qq}.
These predictions show a remarkable agreement with the experimental values.

\section{Discussion about the mixing parameters}
\label{discussion}
In this Section, we compare our best results for the pseudoscalar decay constants and mixing angles
in the octet-singlet basis (third fit of Table \ref{table1}) with the theoretical expectations of 
large $N_c$ ChPT.
These values are extracted from a comparison with experimental data only assuming that
the pseudoscalar decay constants involved in the corresponding processes follow the
parametrization given in Eq.~(\ref{defdecaybasisos}).
Using the experimental constrain $f_K=1.22 f_\pi$, one obtains \cite{Leutwyler:1997yr}:
$f_8=1.28 f_\pi$, $\theta_8=-20.5^\circ$, $f_0\simeq 1.25$, and $\theta_0\simeq -4^\circ$.
As stated, our best fitted results are quite in agreement with the former values
except for the case of $f_8$.
Note however that in Ref.~\cite{Leutwyler:1997yr} the value of $f_8$ is fixed from theory
while in our analysis it is fitted from a direct comparison with experimental data
where a positive correlation between $f_8$ and $\theta_0$ appears.
For $f_8=1.34 f_\pi$ the results of the fit are in clear agreement with the predictions
from large $N_c$ ChPT even though the quality of the fit is slightly reduced.
As seen from Table \ref{table1}, if the constrain $f_8=1.28 f_\pi$ is imposed one gets a worse fit.

Our best results for the mixing parameters
can be used to check the consistency of the set of equations (\ref{mixparthos}),
and therefore to test the reliability on the large $N_c$ ChPT framework.
Accordingly, our fitted values for $f_8$ and $f_0$ together with the third equation in (\ref{mixparthos})
can be used to get $\theta_8-\theta_0=(-13.4\pm 0.7)^\circ$ as a prediction for the difference
of the two mixing angles in the octet-singlet basis.
If one compares this prediction with our result $\theta_8-\theta_0=(-22.7\pm 2.6)^\circ$
a disagreement is again obtained.
Using $f_8=1.34 f_\pi$ and our fitted value for $f_0$ (second fit of Table \ref{table1})
one gets $\theta_8-\theta_0=(-15.8\pm 0.5)^\circ$
to compare with our result $\theta_8-\theta_0=(-18.1\pm 2.6)^\circ$.
The second equation in (\ref{mixparthos}) may be used to get a prediction for the 
OZI-rule violating parameter $\Lambda_1$ once $f_0$ is provided.
Using our value for $f_0$ one gets $\Lambda_1=0.42\pm 0.13$.

The numbers obtained in the former discussion hint at a disagreement with large $N_c$ ChPT.
Some care should be taken however to qualify this statement.
On the one hand, the statistical significance is still limited,
but will obviously improve as critical channels are measured more accurately.
On the other hand, it is plainly clear that the critical information comes here from
radiative decays of vector mesons, where we use vector meson dominance.
Questions have been raised as to the consistency of vector resonances and VMD with ChPT and
short distance QCD cf.~e.g.~\cite{Knecht:2001xc}.
This latter remark should however be moderated by the fact that our approach was carefully tested
in the charged mesons sector (away from the $\eta$-$\eta^\prime$ mixing problems) \cite{Ball:1995zv}.

\section{Summary and conclusions}
We have performed a phenomenological analysis on various decay processes
in order to test the large $N_c$ Chiral Perturbation Theory predictions for the octet and singlet
pseudoscalar decay constants and mixing angles.
First we have derived theoretical expressions for the radiative decays 
$(\eta,\eta^\prime)\to\gamma\gamma$ and for the $VP\gamma$ form factors
using a two mixing angle scheme for the $\eta$-$\eta^\prime$ system.
Second we have used experimental data on these decays and couplings
to fit the values of the mixing parameters in the octet-singlet basis.
Finally, we have compared our best fitted mixing parameters with the expectations from
large $N_c$ ChPT showing a possible discrepancy with this framework.
Higher accuracy data and more refined theoretical analyses would contribute to clarify the
preceding issue.

\noindent
\emph{Acknowledgements:}
I would like to express my gratitude to the PHOTON 2005
Organizing Committee for the opportunity of presenting this contribution,
and for the pleasant and interesting conference we have enjoyed.

\end{document}